# Selection of cyanobacteria over green algae in a photo-sequencing batch bioreactor fed with wastewater

Dulce M. Arias[1], Estel Rueda[1], María J. García-Galán[1], Enrica Uggetti[1]*, Joan García[1]

[1]GEMMA – Group of Environmental Engineering and Microbiology, Department of Civil and Environmental Engineering, Universitat Politècnica de Catalunya•BarcelonaTech, c/ Jordi Girona 1-3, Building D1, E-08034, Barcelona, Spain.

* Corresponding author:

Tel.: +34 934016465

E-mail address: enrica uggetti@upc.edu

**Abbreviations**

| | |
|---|---|
| HRT | Hydraulic retention time |
| Lv | Volumetric load |
| Q | Flow |
| $V_m$ | Mixed liquor discharge volume |
| $V_s$ | Supernatant discharge volume |
| SOC | Soluble organic carbon |
| SRT | Solids retention time |
| TAN | Total ammoniacal nitrogen |
| TIC | Total inorganic carbon |
| TIN | Total inorganic nitrogen |
| TN | Total nitrogen |
| TIP | Total inorganic phosphorus |
| TOC | Total organic carbon |
| TON | Total organic nitrogen |
| TOP | Total organic phosphorus |
| TSS | Total suspended solids |
| $TSS_m$ | Mixed liquor total suspended solids |
| $TSS_x$ | Supernatant total suspended solids |
| TP | Total phosphorus |
| VSS | Volatile suspended solids |

## Abstract

In this work, a strategy based on photo-sequencing batch operation was used to select cyanobacteria over unsettled green algae in a wastewater treatment system, evaluating for the first time the effect of hydraulic regimes on nutritional dynamics and microorganisms' competition. During 30 days of operation, an initial microalgae mixed consortia dominated by the green microalgae *Scenedesmus* sp. was cultivated in two different photo-sequencing batch reactors operated at hydraulic retention time (HRT) of 6 days ($PSBR_6$) and 4 days ($PSBR_4$) at a theoretical solids retention time (SRT) of 10 d. Both reactors were compared with a semi-continuous reactor ($SC_{10}$) operated at 10 d of HRT and 10 days of SRT (used as a control). The results indicated that $PSBR_6$ and $PSBR_4$ decreased *Scenedesmus* sp. population by 88% and 48%, respectively. However, only $PSBR_6$ provided suitable conditions to select cyanobacteria from an initial green algae dominated culture. These conditions included volumetric loads of 11.72 mg TN $L^{-1}$ $d^{-1}$, 2.04 mg TP $L^{-1}$ $d^{-1}$ and 53.31 mg TOC $L^{-1}$ $d^{-1}$. The remaining nutrients in the culture led also to a phosphorus limiting N:P ratio (34:1) that improved the increase of cyanobacteria from an initial 2% until 70% of the total population. In addition, $PSBR_6$ reached a biomass production of 0.12 g $L^{-1}$ $d^{-1}$, while removing TN, TP and TOC by 58%, 83% and 85%, respectively. Conversely, the application of higher nutrients loads caused by lower HRT ($PSBR_4$) led to an increase of only 13% of cyanobacteria while $SC_{10}$ remained with the same biomass composition during all the experimental time. Thus, this study showed that the dominance of cyanobacteria in microalgal-based wastewater treatment systems can be controlled by the operational and nutritional conditions. This knowledge could contribute to control microalgae contamination from up-scaling cyanobacterial biomass production in wastewater treatment systems.

## 1. Introduction

Nowadays, the cultivation and use of microalgae and cyanobacteria as feedstock to obtain biofuels, bioproducts and bioenergy has become a relevant research topic. Increasing scientific interest has been devoted to cyanobacteria (blue-green algae) due to their ability to grow in wastewater effluents and to their capacity to produce and accumulate intracellular bioactive compounds of interest for food and non-food purposes, such as, phycobiliproteins, polyhydroxybutyrates and glycogen (Abed et al., 2009; Shalaby, 2011). These bioproducts can be used as pigments, bioplastics and a biofuel substrates, respectively (Markou et al., 2013; Samantaray and Mallick, 2014; Stal, 1992; Van Den Hende et al., 2016).

The alternative use of wastewater effluents as nutrients source for cyanobacteria growth represents the most promising, cost-effective and eco-friendly strategy to reduce production costs associated with nutrients and water input (Samantaray et al., 2011; Zhou et al., 2012). However, cyanobacteria cultivation in such a variable media implies certain disadvantages related to the competition with other microorganisms, especially with green microalgae. Certain studies have related competition relationships to abiotic factors such as temperature, nutrients and pH (Tang et al., 1997). Although several studies in lakes and reservoirs have dealt with the importance of nutrients' interaction with algal composition (Dolman et al., 2012), there are comparatively few studies focusing on the relation between nutritional conditions in wastewater and the presence of cyanobacteria (Arias et al., 2017; Van Den Hende et al., 2016). Hence, the factors controlling these competition relationships are still not completely understood. Recently, Arias et al. (2017) successfully selected and maintained a dominant population of cyanobacteria from an initial green algae dominated consortium under a long term operation. The dominance of cyanobacteria over green algae as *Chlorella* sp. and *Stigeoclonium* sp. was associated to the total nutrients concentration controlling competence relationships, in particular low inorganic phosphorus loads (<0.23 mg P-$PO_4^{3-}$ L $d^{-1}$) and N:P ratios between 16:1-49:1 (molar basis). However, this factor cannot be considered as the only aspect favoring the competition of cyanobacteria since other green microalgae such as *Scenedesmus* sp. can also tolerate low phosphorus content and high N:P ratios (Arias et al., 2018; Gantar et al., 1991; Xin et al., 2010). Indeed, *Scenedesmus* sp. has been widely reported in different types of wastewater treatment systems (i.e. waste stabilization ponds and high rate algal ponds) because of its high tolerance to a wide range of N and P loads (Xin et al., 2010); besides, it is considered as one of the best competitors for inorganic carbon in comparison to cyanobacteria and other green algae (Ji et al., 2017). These similar optimal nutrient conditions for *Scenedesmus* sp. and cyanobacteria species suggest that the latter could be highly exposed to contamination and competition in mixed cultures. This fact would represent a serious drawback during cyanobacterial biomass production for bioproducts, since *Scenedesmus* sp. is unable to produce certain metabolites that can be accumulated in cyanobacteria (e.g. phicobiliproteins and polyhydroxybutyrates). Moreover, in the case of carbohydrates for biofuels, *Scenedesmus* sp. presents disadvantages due to its hard cellulose cell wall, which typically requires additional pretreatments and further expensive conversion processes to extract the product (Bohutskyi and Bouwer, 2013; Nozzi et al., 2013). Hence, production strategies should be developed in order to improve cyanobacteria competition over these green algae.

Contrary to *Scenedesmus* sp. and other eukaryotic microalgae, many species of cyanobacteria have the ability to easily form aggregates in the culture and promote a high settleability and natural gravity harvesting (Arcila and Buitrón, 2016; de Godos et al., 2014). This advantage could be used as a strategy to select cyanobacteria over unsettling green algae. Thus, the use of optimized hydraulic retention time (HRT) and solids retention time (SRT) during biomass production could be used under continuous operation to select microorganisms able to form flocs fast, while unsettling microorganisms will be continuously removed. However, despite being a promising alternative, uncoupled HRT and SRT can also lead to changes related to nutrients loads and their consequences on microbial changes need to be addressed.

In a previous work by Arias et al., (2018a), it was demonstrated that nutrients dynamics in photo-sequencing batch reactors operated at low HRT and SRT are efficient in the removal of unsettleable microalgae. However, the high carbon and nutrients loads caused by low HRT employed in this study led to high bacteria and low cyanobacteria populations at the end of four weeks of operation. Taking this into consideration, the main objective of this study is to select cyanobacteria from an initial consortium dominated by unsettleable green algae (*Scenedesmus* sp.), by means of applying higher uncoupled HRT and SRT, which represents lower loads of nutrients in a photo-sequencing batch reactor using secondary effluent and digestate as feedstock. This work aims to evaluate the balance in the effect of batch operation (including a settling phase), nutrient loads and ratios, and SRT and HRT conditions on the dominance of cyanobacteria in mixed cultures.

## 2 Material and methods

### 2.1 Experimental set up

#### *2.1.1 Inoculum*

The experimental set-up was located at the laboratory of the GEMMA research group (Universitat Politècnica de Catalunya. BarcelonaTech, Spain). A mixed culture was used as inoculum. Green microalgae, cyanobacteria, diatoms, protozoa and rotifers abundance was evaluated by microbial counting and revealed a consortium mostly dominated by *Scenedesmus sp.* (93±2%), with other species of green algae (4±1%), cyanobacteria (2±1%) and diatoms (1±0.01%). Microscopic images of the initial culture biomass are shown in Fig. 1. It was obtained from a pilot tertiary wastewater treatment system consisting of a closed photobioreactor (30 L) fed with urban secondary wastewater and liquid digestate (not centrifuged), operated at a HRT of 10 days. Operational details and other characteristics of this system can be found in detail elsewhere (Arias et al., 2018a). The biomass was collected from a harvesting tank connected to the photobioreactor and it was thickened by gravity in laboratory Imhoff cones during 30 min before inoculation.

*2.1.2 Experimental set-up*

The experiments were carried out indoors and performed with three experimental closed photobioreactors of polymethacrylate and cylindric shape, with an inner diameter of 11 cm, a total volume of 3 L and a working volume of 2.5 L (Fig. A1).

All the photobioreactors were continuously maintained in alternate light:dark phases of 12 h and continuously stirred (with the exception of the settling periods) with a magnetic stirrer (selecta, spain) set at 250 rpm to ensure a complete mixing in the photobioreactor. Temperature was continuously measured by a probe inserted in the photobioreactor (ABRA, Canada) and kept approximately constant at 27 (±2) °C by means of a water jacket around the photobioreactor. Illumination during the light phase was supplied by two external halogen lamp (60W) placed 11 cm from each photobioreactor. Each two sets provided a constant 220 μmol $m^{-2}$ $s^{-1}$ of light. pH was continuously measured with a pH sensor (HI1001, HANNA, USA) and kept at 8.5 with a pH controller (HI 8711, HANNA, USA) by the automated addition of HCl 0.1 N and NaOH 0.1 N. This pH set point of 8.5 was selected based on previous literature that reported a pH preference of cyanobacteria ranging from 8 to 9 (Ahn et al., 2002; Arias et al., 2017; Reynolds, 1987; Unrein et al., 2010). Automatic addition and discharge of the mixed liquor, the supernatant and the feeding in the photobioreactors was carried out automatically by peristaltic pumps.

Two photo-sequencing batch reactors were operated with different uncoupled hydraulic retention time (HRT) and solids retention time (SRT) during an experimental time of 30 days. As observed in our previous publication (Arias et al., 2018a), a HRT of 2 days led to increase eukaryotic microalgae and bacterial populations, and for this reason in the present study higher HRT (4 and 6 days) were tested in order to reduce the carbon and nutrients loads in the culture. Theoretical SRT in both photobioreactors was fixed at 10 days. This SRT was chosen based on the results of the previous works (Arias et al., 2017, 2018a, 2018b), demonstrating that SRT lower than 10 affects negatively cyanobacterial growth.

A semi-continuous photobioreactor (namely $SC_{10}$), operated with coupled HRT/SRT of 10 days was used as control in order to evaluate the effectiveness of uncoupling those operational conditions in relation to the removal of unsettleable green algae.

*2.1.2.1 Photo-sequencing batch reactors and semi-continuous reactor set-up*

Each photobioreactor was inoculated with 100 mL of thickened biomass and filled up to 2.5 L with deionized water. From the first day of operation the volume corresponding to the HRT chosen for each photobioreactor was removed, meaning that deionized water was almost completely removed after the first HRT.

One photo-sequencing batch reactor, named $PSBR_6$, was a photobioreactor operated with an HRT of 6 days, while the other, named $PSBR_4$, was operated at an HRT of 4 days. The operational diagram of each photobioreactor can be observed in Fig. 2.

In $PSBR_6$, 0.250 L of the mixed liquor were discharged at the end of the dark phase. Later, the agitation was stopped and biomass was allowed to settle during 30 minutes, followed by the discharged of 0.167 L of the supernatant. Total volume removed was therefore of 0.417 mL, which was replaced by the same volume of wastewater influent.

In $PSBR_4$, 0.250 L of the mixed liquor were discharged at the end of the dark phase, followed by 30 min of biomass settling and a subsequent discharge of 0.375 L of the supernatant. Total volume removed of 0.625 L was replaced by the same volume of wastewater influent. It should be noticed that real SRT calculated in the photo-sequencing batch reactors was conditioned by the solids discharged in the mixed liquor prior to sedimentation and the solids contained in the supernatant.

On the other hand, the semi-continuous reactor ($SC_{10}$) was operated by removing 0.250 L of the mixed liquor at the end of the dark phase and subsequently replaced by the same volume of wastewater influent.

*2.1.2.2 Wastewater*

Wastewater influent consisted in digestate diluted in secondary effluent in a ratio of 1:50. Digestate was obtained daily from a lab-scale microalgae anaerobic digester (1.5 L, flow of 0.075 L $d^{-1}$), operated at 35 °C and a SRT and HRT of 20 days. Secondary effluent was obtained from a secondary settler after a high rate algal pond (HRAP) treating urban wastewater (Gutiérrez et al., 2016a; Passos et al., 2014). Dilution of 1:50 was chosen according to a previous study by Arias et al., (2017), which showed an annual nutritional concentration <15 mg total ammoniacal nitrogen (TAN) $L^{-1}$, preventing TAN toxicity, avoiding high turbidity in the feeding and allowing suitability for cyanobacteria growth.

Characteristics and operation of the anaerobic digester are detailed in Passos et al. (2013). Characteristics of the digestate, the secondary effluent and the wastewater influent mixture are shown in Table 1.

*2.3. Microalgal population*

Quantitative analysis of green microalgae, cyanobacteria, diatoms, protozoans and rotifers was performed once a week by microscopic area counting (cells $mL^{-1}$) (Guillard and Sieracki, 2005). To this aim, 20 µL of mixed liquor were added to a slide with a coverslip and individual cells were counted per field until reach >400 cells to have a total standard error lower than 5% (Margalef, 1983) alternating bright field microscopy and fluorescence microscopy. Microalgae, diatoms, protozoa and rotifers were quantified in bright field microscopy at 40X, while cyanobacteria species were counted using fluorescence microscopy at 40X with the operation of filters containing an excitation filter (510-560 nm), emission filter (590 nm) and dichroic beam splitter (575 nm). Both bright field and fluorescence microscopy were performed using a fluorescence microscope (Eclipse E200, Nikon, Japan).

Qualitative evaluation of microalgae composition changes within each photobioreactor was monitored by microscopy once a week. Microbial visualization was performed in bright field microscopy at 40X in an optic microscope (Motic, China) equipped with a camera (Fi2, Nikon, Japan) connected to a computer (software NIS-Element viewer®). Cyanobacteria and microalgae species were identified *in vivo* using conventional taxonomic books (Bourrelly, 1985; Palmer, 1962), as well as a database of cyanobacteria genus (Komárek and Hauer, 2013).

*2.4 Analytical methods*

Samples taken from influent wastewater (digestate + secondary effluent) and effluent of the three photobioreactors (supernatant after settling) were analyzed for nutrients concentration. It should be noticed that dissolved nutrients in the effluent are equivalent to the nutrients contained in the mixed liquor of the culture. Note that in the case of $SC_{10}$, the supernatant sample was taken from the mixed liquor discharged and subsequently submitted to a separation process in laboratory Imhoff cones during 30 min before in order to remove biomass from the effluent.

Nutrients analysis were performed three times per week for total organic carbon (TOC), total inorganic carbon (TIC) and soluble organic carbon (SOC), total inorganic phosphorus (TIP, P-$PO_4^{3-}$), nitrite (N-$NO_2^-$), nitrate (N-$NO_3^-$) and total ammoniacal nitrogen (TAN, sum of N-$NH_3^-$ and N-$NH_4^+$). Total nitrogen (TN) and total phosphorus (TP) were measured two days per week. TAN was determined using the colorimetric method indicated in Solorzano (1969). Ion chromatography was used to measure concentrations of IP, N-$NO_2^-$ and N-$NO_3^-$ by means of a DIONEX ICS1000 (Thermo-scientific, USA), whereas TOC, TIC and TN were analyzed by using a C/N analyzer (21005, Analytikjena, Germany). TP was analyzed following the methodology described in Standard Methods (APHA-AWWA-WPCF, 2001). Total inorganic nitrogen (TIN) was calculated as the sum of N-$NO_2^-$, N-$NO_3^-$ and TAN. Total organic nitrogen (TON) (the sum of dissolved and particulate form) was calculated as the difference between TN and N-$NO_2^-$ and N-$NO_3^-$, whereas total organic phosphorus (TOP) (the sum of dissolved and particulate form) was determined as the difference between TP and TIP.

Total suspended solids (TSS) and volatile suspended solids (VSS) were measured in the mixed liquor (in all photobioreactors) and in the supernatant (only in $PSBR_6$ and $PSBR_4$) three days per week, while Chlorophyll *a* concentration was measured in the cultures once per week. Those procedures were based in the methodology described in Standard Methods (APHA-AWWA-WPCF, 2001). All parameters defined above were determined in triplicate, and measured in samples taken at the end of the dark phase.

*2.5. General calculations*

Actual SRT [$d^{-1}$] was calculated as follows:

$$\mathbf{SRT} = \frac{V}{[Vm+Vs(\frac{TSSx}{TSSm})]} \qquad (1)$$

Where: V [$L^{-1}$] is the total volume of the photobioreactor, $V_m$ [$L^{-1}$] is the mixed liquor discharge volume, $V_s$ is the supernatant discharge volume, $TSS_m$ [mg $L^{-1}$] is the mixed liquor suspended solids concentration and $TSS_x$ [mg $L^{-1}$] is the supernatant suspended solids concentration.

Settleability [%] was determinate according to the following equation:

$$\mathbf{Settleability} = \mathbf{100} * [\mathbf{1} - \left(\frac{\boldsymbol{TSSs}}{\boldsymbol{TSSx}}\right)] \qquad (2)$$

Where $TSS_m$ [mg $L^{-1}$] is the mixed liquor suspended solids concentration and $TSS_x$ [mg $L^{-1}$] is the supernatant suspended solids concentration.

Biomass production of each photobioreactor in [g VSS $L^{-1}$ $d^{-1}$] was estimated using the equation:

$$\mathbf{Biomass\ production} = \frac{\boldsymbol{Q*VSS}}{\boldsymbol{V}} \qquad (3)$$

where Q is the flow [$L^{-1}$ $d^{-1}$], VSS is the biomass concentration in the photobioreactor [g $L^{-1}$] and V [$L^{-1}$] is the volume of the photobioreactor.

Applied nutrients (TOC, TIC, TAN, N-$NO_2^-$, N-$NO_3^-$, TIN, TON, TN, TIP, TOP and TP) volumetric load (Lv-X) in [mg X $L^{-1}d^{-1}$] was calculated as follows:

$$\mathbf{Lv - X} = \frac{Q*X}{V} \quad (4)$$

Where Q is the flow [$L^{-1}$ $d^{-1}$], X is the corresponding nutrient influent concentration [mg X $L^{-1}$] and V [$L^{-1}$] is the volume of the photobioreactor.

## 3. Results

### *3.1 Settling efficiencies and actual solids retention time (SRT)*

In this study the presence of a settling phase to the photobioreactors operation was assessed in order to select flocs forming biomass and remove the unsettled biomass, composed mostly by green algae *Scenedesmus* sp., achieving an increase in the abundance of cyanobacteria from an initial mixed green algae-cyanobacteria consortium. Indeed, the settling phase tested promoted cyanobacteria dominance, but the relative abundance of cyanobacteria in the different photo-sequencing batch reactors was also influenced by the SRT and nutritional loads.

As observed in Fig. 3, settleability increased from 38% to 90% in $PSBR_6$ and from 24% to 77% in $PSBR_4$. This increase was caused by the natural gravity harvesting of cultures. These changes in settleability affected the SRT. In $PSBR_6$, the actual SRT at day 1 was 7.1 d, increasing until reaching a SRT of 9 days, which remained constant after day 12 until the end of the experiment. On the other hand, the operation of $PSBR_4$ started with an actual SRT of 6.6 d that increased until reaching a SRT of approximately 8 d after the day 8 of operation, which maintained quite constant from this day.

*3.2 Biomass production and microbial evolution*

Changes in settleability and actual SRT also had an impact in biomass concentration, as shown in Fig. 4. Biomass concentration in $SC_{10}$ (with a hydraulic retention time (HRT)/SRT of 10 d) showed a quite constant pattern during all the experimental time (0.56±0.05 g VSS $L^{-1}$). $PSBR_6$ and $PSBR_4$ showed an increase in the biomass during the first 12 days of operation, until reaching a steady state from day 15 onwards 0.57±0.09 g VSS $L^{-1}$ and 0.65±0.072 g VSS $L^{-1}$, respectively. The highest biomass production was achieved in $PSBR_4$ (0.16 g VSS $L^{-1}$ $d^{-1}$), higher than that observed in $SC_{10}$ (0.06 g VSS $L^{-1}$ $d^{-1}$) and $PSBR_6$ (0.12 g VSS $L^{-1}$ $d^{-1}$).

Microscopic qualitative observations during the last 10 days of operation, when all the reactors showed a biomass steady state, are provided in Fig. 5 and clearly show how the culture in the $SC_{10}$ remained mostly dispersed and with a population dominated by *Scenedesmus* sp. On the contrary, the cultures in $PSBR_6$ were turned into a culture dominated by flocs throughout the experimental time. More interestingly, a more predominant presence of flocs formed by cyanobacteria cf. *Aphanocapsa* sp. were observed. In the case of $PSBR_4$, dispersed cells of *Scenedesmus* sp. were observed in the culture, and other green algae such as *Stigeoclonium* sp. were also present.

These observations agreed with the microscope quantitative counting of the biomass in the three photobioreactors. Biomass in $SC_{10}$ showed the dominance of *Scenedesmus* sp. throughout the experimental time with a slow decreasing trend, from an initial abundance of 91% to 74%. While other species such as cyanobacteria and green algae as *Chlorella* sp. and *Stigeoclonium* sp. increased from the initial 4% to 7% and 14%, respectively (Fig. 6a). In this case, diatoms and protozoa remained with the same abundance than in the beginning (1-2%) during all the experimental time.

On the other hand, the initial *Scenedesmus* sp. abundance in $PSBR_6$ was reduced from 95% to only 7% (Fig. 6b). Interestingly, *Scenedesmus* sp. population lowered while cyanobacteria abundance (mostly composed by cf. *Aphanocapsa* sp.) increased from 2% until up to 70% from day 9 onwards. The abundance of other green algae different than *Scenedesmus* sp. also gradually increased from 3% to 28% on day 19 and then decrease to 21% on day 33. Other microorganisms as diatoms and protozoa remained low (2% and 1%, respectively) during all the experimental time.

In $PSBR_4$, *Scenedesmus* sp. abundance was gradually reduced from the initial 94% until reaching 52%. On the contrary, cyanobacteria abundance increased, reaching 15% in the last day of operation from an initial 2%. Other green algae abundance also increased from 4% until 29% in the last day of operation (Fig. 6c). Other microorganisms as diatoms and protozoa where also observed in all the experimental time, remaining an abundance of 1-2%.

Although the presence of heterotrophic bacteria was not directly quantified in this study, it was indirectly evaluated through the ratio of chlorophyll *a* mass (implying green algae and cyanobacteria) divided by total biomass (VSS) (which integrates the biomass of all microorganisms). Thus, the highest content of photosynthetic microorganism was found in $SC_{10}$ control, with a Chlorophyll *a*/VSS ratio of 1.03 mg Chl *a*/g VSS, whereas $PSBR_6$ and $PSBR_4$ reached values of 0.56 and 0.53 mg Chl *a*/g VSS, respectively. Lower values in photo-sequencing batch reactors indicate higher content of bacteria, also according to the higher organic applied.

*3.3 Nutritional conditions*

*3.3.1 Nutrients uptake and wastewater treatment*

Unsettling and settling operation performed in this study led to different nutritional loads in the photobioreactors by means of the operational HRT (Table 2). In general, the volumetric loads Lv-TIC/Lv-TOC, as well as all the nitrogen and phosphorus forms, increased in each photobioreactor in relation to the decrease in HRT. These differences in the loads caused different culture compositions (Table 3) and led a selective pressure on microbial populations, as has been shown in previous Sections.

Total nitrogen, phosphorus and carbon concentrations in the effluent (therefore, concentrations in the photobioreactors) with respect to the load applied are shown in Fig. 7. As it can be observed, TN and TP were completely consumed in $SC_{10}$ during the experiment (Fig. 7a). TIN was almost completely eliminated during the experiment, whereas TON decreased gradually. Conversely, TIP was completely removed during the experimental time. Likewise, TOP was completely removed along the experiment.

In the case of the effluent of $PSBR_6$, TN was not completely removed due to an increase in TIN (Fig. 7b). It should be noticed that the Lv-TIN was composed of 60% TAN and 30% $N\text{-}NO_2^-$+$N\text{-}NO_3^-$ (Table 1), and in the effluent, 98% of the TIN concentration was based on $N\text{-}NO_3^-$ values. On the other hand, TIP was completely consumed during the first 15 days of operation, and afterwards it occasionally showed effluent concentrations above 1.5 mg $L^{-1}$. This increase in the TIP concentration can be associated to the mineralization of organic phosphorus, since TOP concentration in the effluent maintained values between 3.02 to 4.2 mg $L^{-1}$, which means that the process of mineralization of organic phosphorus to inorganic phosphorus was sometimes slow and the microorganisms didn't have enough time to uptake the TIP (Table 3). Note that a fraction of TON and TOP concentration is related to particulate species. Thus, concentrations of TOP may be associated to intracellular polyphosphates accumulated in unsettling biomass.

$PSBR_4$ effluent showed a similar pattern to that found in $PSBR_6$ regarding TIN, TON, TIP and TOP concentrations. However, $PSBR_4$ showed a better TN removal despite the higher loads applied. Hence, a complete transformation of TON to TAN was observed along the experiment. While an efficient removal of the influent TIN was achieved during the first half of the experiment (Fig. 7c). After that period, TIN values remained around 10 mg $L^{-1}$ (mostly in the form of N-$NO_3^-$), that are lower than those observed in $PSBR_6$. Similarly, TIP was also completely removed during the first half of the experiment and afterwards the effluent showed values around 2 mg $L^{-1}$, mostly due to the mineralization of the accumulated TOP, that was more evident in the second half of the experiment (Fig. 7c).

On the other hand, according to the SOC concentrations in the photobioreactors, a better removal was obtained with high loads. For instance, $SC_{10}$ showed higher concentrations (61.8±12.0 mg $L^{-1}$) than $PSBR_6$ (48.8±8.0 mg $L^{-1}$), indicating a better elimination of organic matter in spite of having initial higher loads than that of the $SC_{10}$ (Fig. 7b). In the case of $PSBR_4$, it showed similar concentrations than $PSBR_6$. As observed in Figure 7c, with the exception of the results obtained on day 4, the concentrations of TOC along the experimental time showed values of approximately 50 mg $L^{-1}$.

Main nutrients and organic matter removals are presented in Table 4. In general, $SC_{10}$ showed the highest percentage removal of all N and P forms (>80%), whereas $PSBR_6$ and $PSBR_4$ reached values ranging from 54% to 78% of TN as well as TP removal. Indeed, previous studies have related high nutrients removal efficiencies in photobioreactors operated under longer HRT (Muñoz and Guieysse, 2006). Remarkably, in spite that $PSBR_6$ and $PSBR_4$ showed lower percentage removal, they reached higher removal rates than $SC_{10}$ due to the higher volume of wastewater treated.

Conversely, in the case of organic matter removal, $SC_{10}$ showed the lower removal efficiency, achieving only 41% of TOC removal, while $PSBR_6$ and $PSBR_4$ doubled that removal percentage (Table 4). Similarly, higher removal rates were observed in $PSBR_6$ and $PSBR_4$ with respect to $SC_{10}$, more specifically $PSBR_4$ showed the highest removal rates of all the photobioreactors. This lower performance of $SC_{10}$ in comparison with both photo-sequencing batch reactors can be associated to higher microalgae content, releasing organic carbon compounds and thus contributing to the dissolved carbon in the culture.

On the other hand, TIC, which is consumed by autotrophic microorganisms, was unlimited in the three photobioreactors. In the case of $SC_{10}$, influent TIC concentration was removed only by 43% along the experiment, while $PSBR_6$ showed an average percentage removal of 55% and the highest removal rate and $PSBR_4$, presented the lowest TIC removal by 13% (Table 4).

**4. Discussion**

According to the results obtained, the competition dynamics of *Scenedesmus* sp., other green algae and cyanobacteria in $SC_{10}$, $PSBR_6$ and $PSBR_4$ were defined by the pressure created by coupling and uncoupling HRT/SRT. It is clear that the $SC_{10}$, employing a coupled SRT/HRT, maintained the same microalgaepopulation during most of the experiment. However, in $PSBR_6$ and $PSBR_4$ the settling phase which allowed uncoupling of HRT and SRT affected the competition dynamics of the microorganisms present in the culture. It is interesting to note that $PSBR_6$ showed a higher efficiency in removing unsettled *Scenedesmus* sp. than $PSBR_4$, despite that the latter had a highest unsettled volume discharged. This implies that other factors such as organic loading, nutrients content, and changes in SRT than the washing-out of unsettled microorganisms provided by the uncoupled SRT/HRT also played an important role in microorganism's dynamics.

Biomass composition and growth presented different patterns according to the N and P concentrations and ratios in the influent and effluent of each photobioreactor. Even though influent TN:TP ratio was the same for all the photobioreactors (7.9:1), nutrient loads were different (Table 2) and led to different effluent N:P ratios, biomass production and the main dominant algae (Table 5). It should be noticed that in the present study calculated influent N:P ratio in the influent includes all dissolved and particulate forms, whereas effluent ratio only includes TIN:TIP ratio, since it represents the direct available nutrients for microorganisms (Pick and Lean, 1987). In $SC_{10}$, the dominance of *Scenedesmus* sp. over other species can be associated to the low loads of N and P introduced to the culture (Table 2) in addition to the absence of a settling phase. Low nutrient loads promoted nitrogen limitation in the culture (low TIN:TIP ratio).

The ratio obtained in $PSBR_6$ is within the optimum range of 16:1 - 49:1, proposed in a previous study by Arias et al., (2017) and that related the dominance of wastewater borne cyanobacteria in competition to green algae to high TIN:TIP values. This high TIN:TIP ratio, suggesting P limitation in the culture favoured the increase in cyanobacteria population, in particular the dominance of unicellular cf. *Aphanocapsa* sp.. In addition, the washing-out of unsettling microorganisms such as *Scenedesmus* sp. contributed to improve the dominance of cyanobacteria. It should be noted that, despite of P limitation, cyanobacteria dominated culture in $PSBR_6$ achieved a higher biomass production than $SC_{10}$. This fact can be associated to the lower HRT of $PSBR_6$ and also to the capability of cyanobacteria to accumulate phosphorus as polyphosphates and perform luxury uptake (uptake of P in excess of their need for growth) (Cottingham et al., 2015). High biomass production of cyanobacteria under limiting P conditions has also been observed in previous studies (Arias et al., 2017; Arias et al., 2018).

In the case of $PSBR_4$, high nutrients loads (Table 2) led to a TIN:TIP ratio in the influent similar to the Redfield ratio (16:1 in molar bases), considered the optimum ratio for microalgae growth (Redfield, 1958). This caused a fast growth and reestablishment of *Scenedesmus* sp., and in turn a lower settleability in $PSBR_4$ than in $PSBR_6$ (Fig. 3). Furthermore, the high Lv-TOC added to this photobioreactor also improved the heterotrophic bacterial activity, contributing to a higher biomass production (Table 5).

Another factor that should be considered regarding the dominance of cyanobacteria in $PSBR_6$ is the calculated SRT, which was always above 9 days, and in consequence slow growing microorganisms such as cyanobacteria were favoured. On the contrary, calculated SRT of $PSBR_4$ led to an HRT close to 8 days, which could have influenced the predominance of microorganisms able to perform a faster growth rate. In this case, cells of *Scenedesmus* sp. and other green algae that were unsettled could have been able of reestablish (at least, partially) the cell concentration that was daily retired from the supernatant. This same trend was observed in previous studies, in which either cyanobacteria or green algae dominance was enhanced depending on the SRT and under similar nutritional conditions (e.g. cyanobacteria in SRT of 10 d (Arias et al., 2017; Hu et al., 2017) and *Scenedesmus* sp. in 8 d (Arias et al., 2018; Gutiérrez et al., 2016; Passos et al., 2014)). It should be highlighted that the lower biomass production of $SC_{10}$ and $PSBR_6$ could also be related to the higher SRT registered (9-10 days); such high values are associated to a slow growth rate in the culture (de Godos et al., 2014; Valigore et al., 2012).

Comparing all these results with our previous study (Arias et al., 2018a), in which the photo-sequencing batch reactors were operated at HRT of 2, negatively affected cyanobacteria competition in the culture. High nutrients (37.60 mg TN $L^{-1}$ $d^{-1}$ and 5.63 mg TP $L^{-1}$ $d^{-1}$) and high carbon loads (186.10 mg TC $L^{-1}$ $d^{-1}$) provided by the low HRT increased bacterial activity instead and also the appearance of filamentous green algae, diatoms, rotifers and protozoa. The encompassing of this previous work with the results of the present study allows concluding that sequencing batch operation can be used to select cyanobacteria from a green algae dominant culture. According to our overall results, the conditions for improving cyanobacteria dominance include the use of closed stirred PSBR fed with secondary effluent, operated at an HRT of 6 d and a SRT of 10 d, and with loads of approximately of 11.72 mg TN $L^{-1}$ $d^{-1}$, 2.04 mg TP $L^{-1}$ $d^{-1}$, 53.31 mg TOC $L^{-1}$ $d^{-1}$ and 18.5 mg TIC $L^{-1}$ $d^{-1}$. Under these conditions, the culture would be able to have residual nutrients concentrations which allow for a phosphorus limiting N:P ratio in the culture (34:1) and improve the increase of the total cyanobacteria populations from an initial 1% until 70%. Additionally, this operation led to a biomass production up to an average of 0.12 g $L^{-1}$ $d^{-1}$ while removing 3.85 mg $L^{-1}$ $d^{-1}$ of TN, 1.81 mg $L^{-1}$ $d^{-1}$ of TP and 55.38 mg $L^{-1}$ $d^{-1}$ of TC. On the contrary, the decrease of HRT favoured the presence of green algae, diatoms and heterotrophic bacteria.

To the authors knowledge, the strategy of uncoupling SRT and HRT has been employed mostly to promote the formation of green-algae and cyanobacteria aggregates and the enhancement of nutrients rates removal, as shown in the studies of Van Den Hende et al., (2014; 2016), and Wang et al., (2015). However, in these studies, the effect of operational conditions on nutritional dynamics and the competition of specific microorganisms were not considered. Thus, this work addresses this issue for the first time, as well as the influence of factors such as nutrients loads, settleability and microbial evolution in the selection process. Such knowledge about the influence of these factors in microorganism’s competition could facilitate the cultivation of cyanobacteria controlling contamination by other species (even in pure cultures). In this way, cyanobacteria cultivation could be integrated into a real wastewater treatment plant in order to treat wastewater treatment effluents, while producing valuable products used for non-food purposes such as, phycobiliproteins, polyhydroxybutyrate and glycogen (Abed et al., 2009; Shalaby, 2011). Following the encouraging results obtained from this study, it is worthwhile to investigate whether the operational conditions tested could influence the accumulation of those byproducts. Further research addressing these issues need to be done before scaling-up of the technology.

## 5. Conclusions

In this study, it has been proven that cyanobacteria could be selected from an initial consortium dominated by unsettleable green algae (*Scenedesmus* sp.), by means of applying uncoupled HRT and SRT in a photo-sequencing batch reactor using secondary effluent and digestate as feedstock. The results indicated that a photo-sequencing batch reactor operated at 6 days of HRT provided suitable conditions to select cyanobacteria from a green algae dominant culture. These conditions included volumetric loads of 11.72 mg TN $L^{-1}$ $d^{-1}$, 2.04 mg TP $L^{-1}d^{-1}$ and 71.81 mg TC $L^{-1}$ $d^{-1}$. The remaining nutrients in the culture led also to a phosphorus limiting N:P ratio in the culture (34:1). Altogether the conditions of this reactor allowed to increase cyanobacteria population from an initial percentage of 2% to 70% at the end of the experiment.

## Acknowledgments

The authors would like to thank the European Commission [INCOVER, GA 689242] and the Spanish Ministry of Science and Innovation [project FOTOBIOGAS CTQ2014-57293-C3-3-R] for their financial support. Dulce Arias kindly acknowledge her PhD scholarship funded by the National Council for Science and Technology (CONACYT) [328365]. M.J. García and E. Uggetti would like to thank the Spanish Ministry of Industry and Economy for their research grants [FJCI-2014-22767 and IJCI-2014-21594, respectively]. Authors appreciate Nicolò Brena and Laura Torres for their contribution during the experiment deployment.

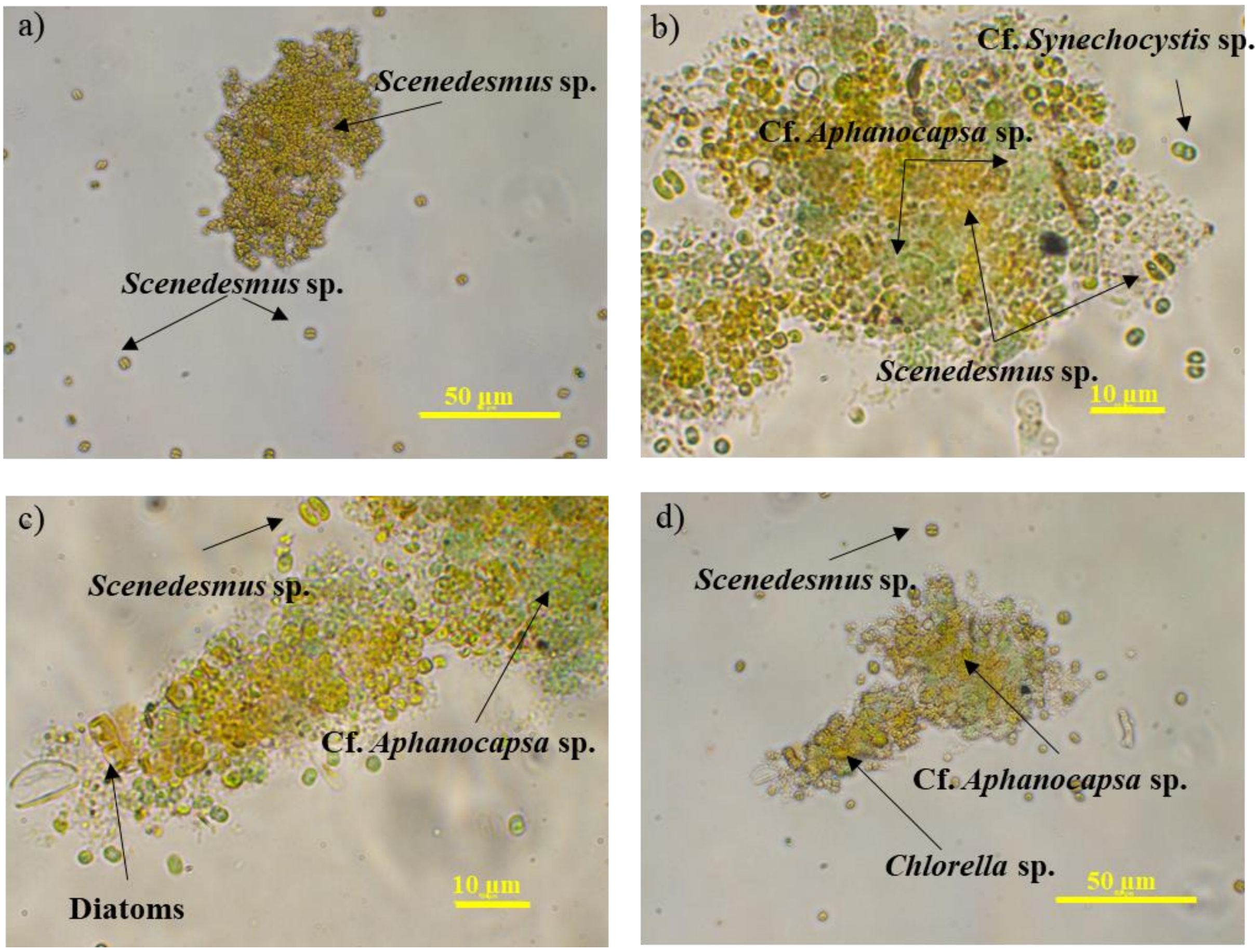

Fig. 1. Microscopic images of the inoculum observed in bright light microscopy at a) 400x, b), c) 1000x and d) 400x. Algal flocs and dispersed cells are composed of green algae *Scenedesmus* sp. and *Chlorella* sp., cyanobacteria cf. *Aphanocapsa* sp., cf. *Chroococcus* sp. and diatoms.

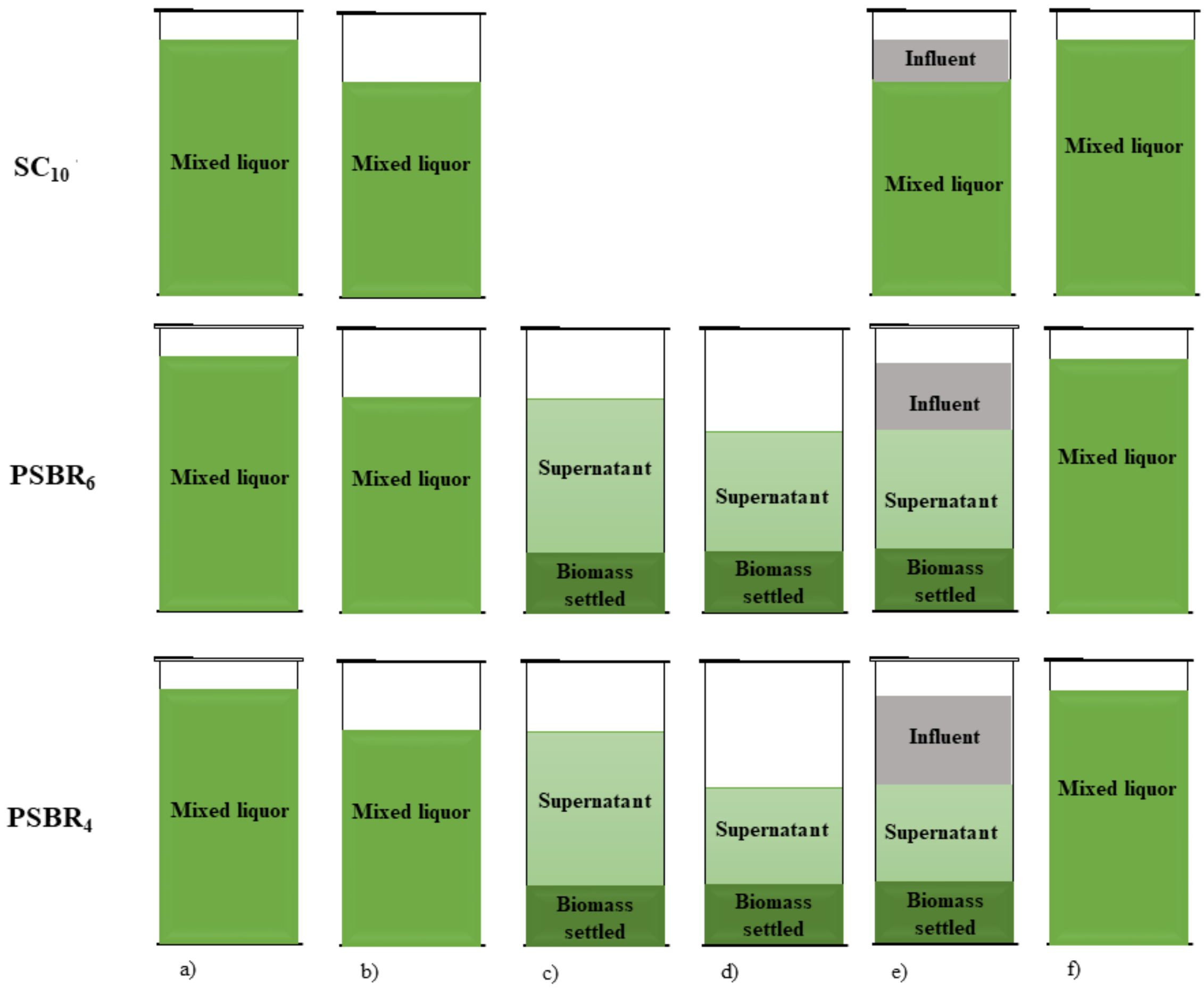

Fig. 2. Scheme of operation of the photobioreactors showing the process during the last minutes of the dark phase; a)-b) biomass discharge (5 min); c) biomass settling (30 min); d) supernatant discharge (10 min); e-f) effluent addition (10 min) and mixing. Biomass separation in $SC_{10}$ was performed in an independent process.

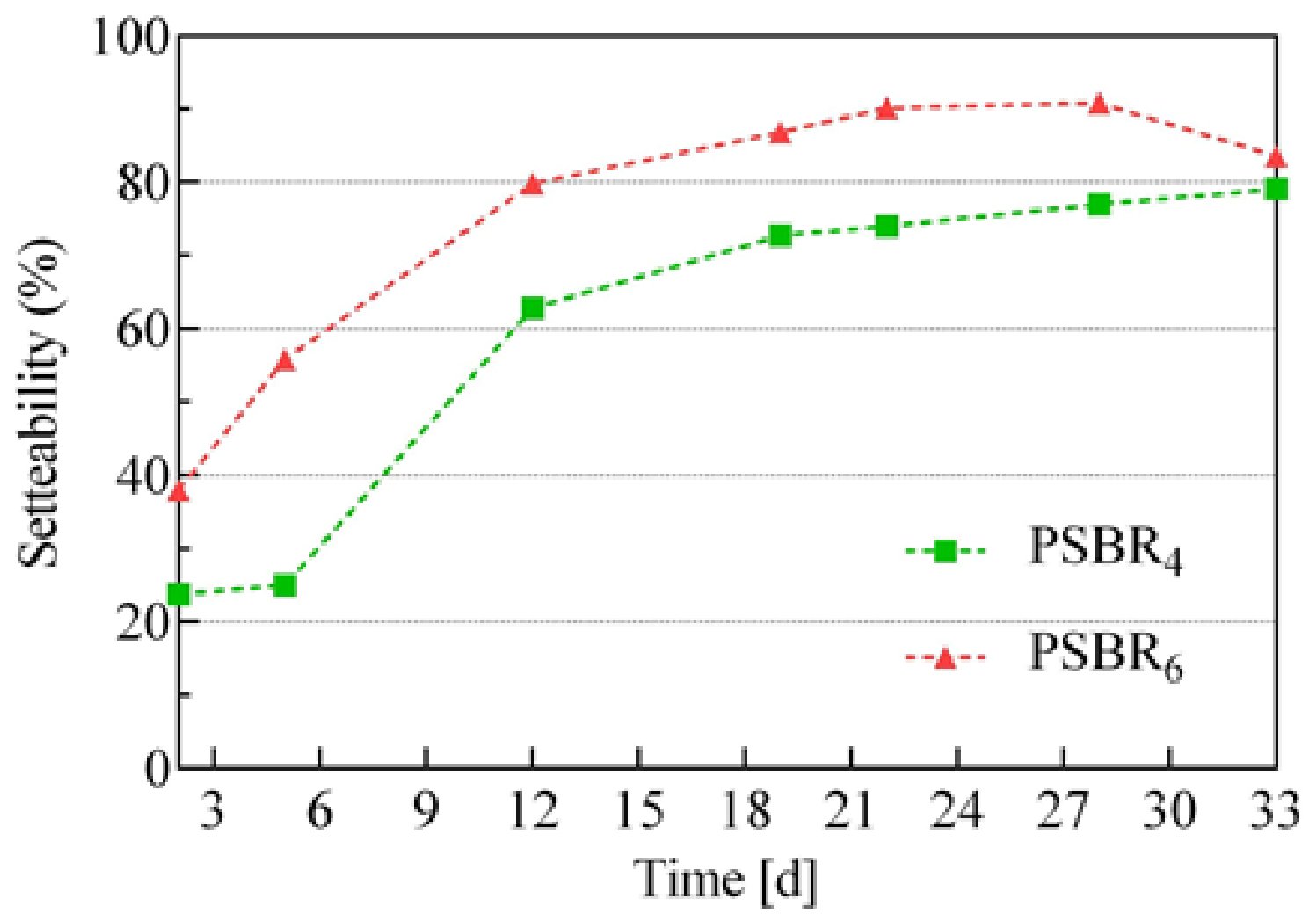

Fig. 3. Settleability after 30 min along the experimental time obtained in $PSBR_6$ and $PSBR_4$ after 30 minutes of settling.

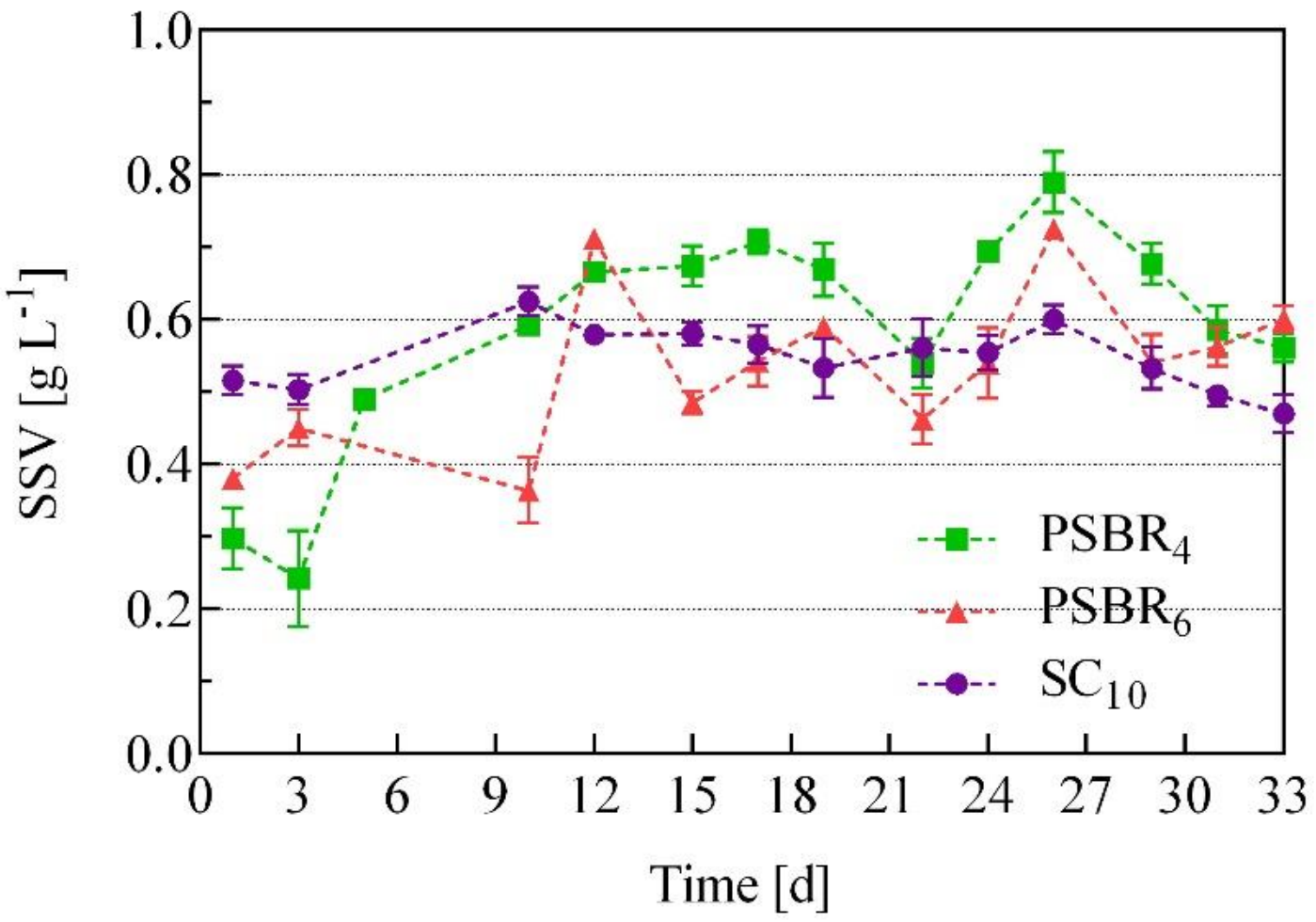

Fig. 4. Biomass concentration changes in different photobioreactors $SC_{10}$, $PSBR_6$ and $PSBR_4$) during the experimental time. Biomass is given as volatile suspended solids (g VSS $L^{-1}$).

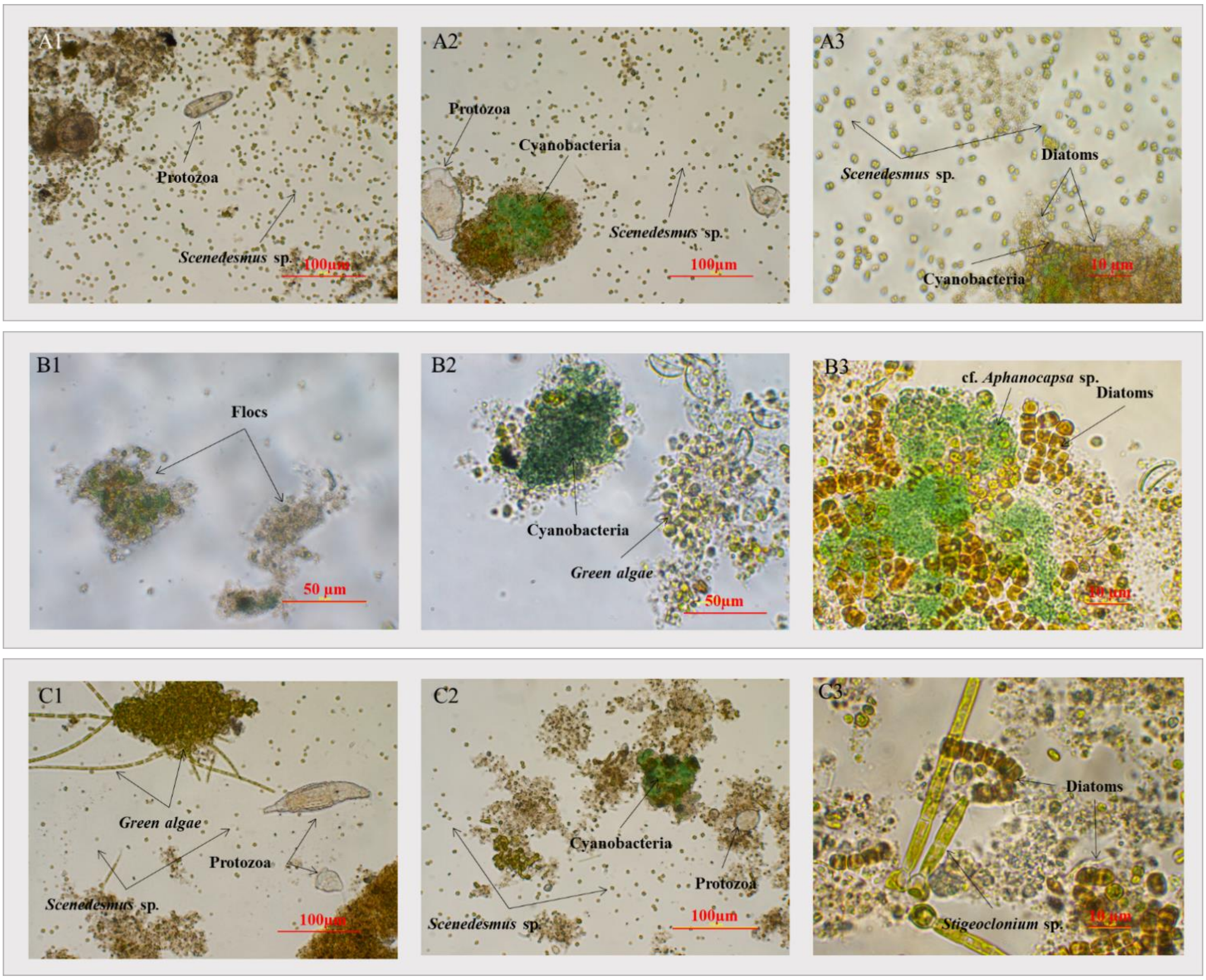

Fig. 5. Microscopic images illustrating the microbial composition during the last days of operation of A) $SC_{10}$, operated with a HRT of 10 d, observed at 200x (A1, A2) and 1000x (A3); B) $PSBR_6$, operated with a HRT of 6 d observed at 400x (B1, B2) and 1000x (B3) and C) $PSBR_4$ operated with a HRT of 4 d, observed at 200x (C1, C2) and 1000x (C3). All the images were observed in bright light microscopy.

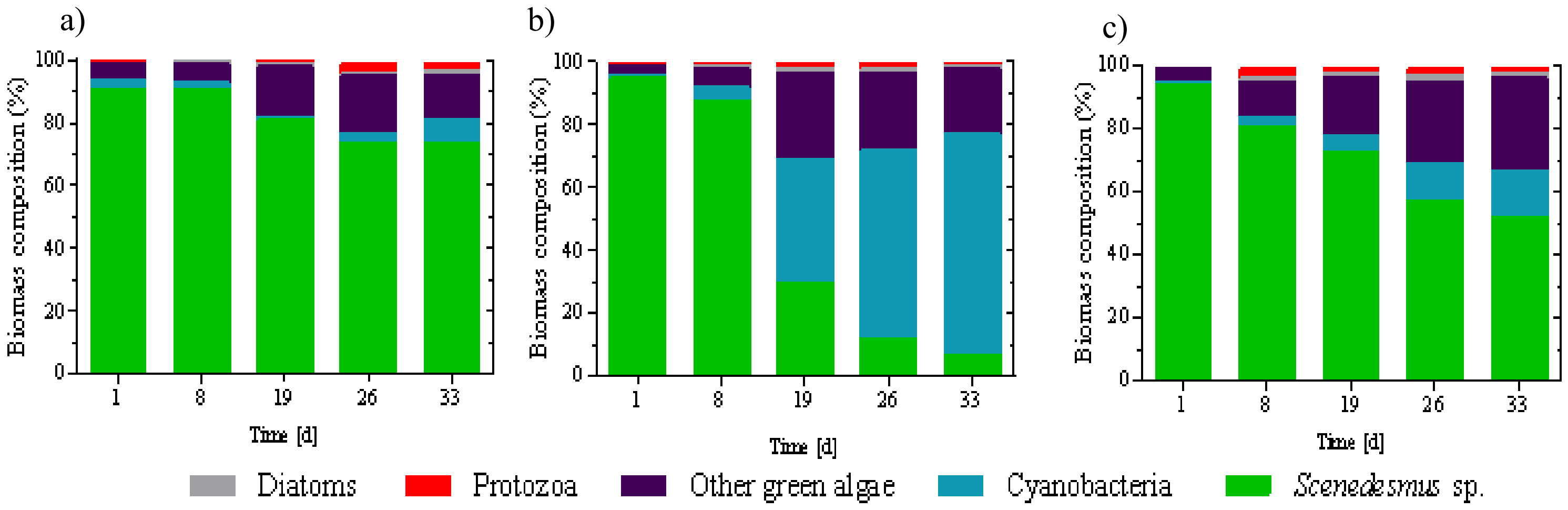

Fig. 6. Biomass composition of different photobioreactors a) $SC_{10}$, b) $PSBR_6$ and c) $PSBR_4$. Percentages were calculated considering the total cells $mL^{-1}$.

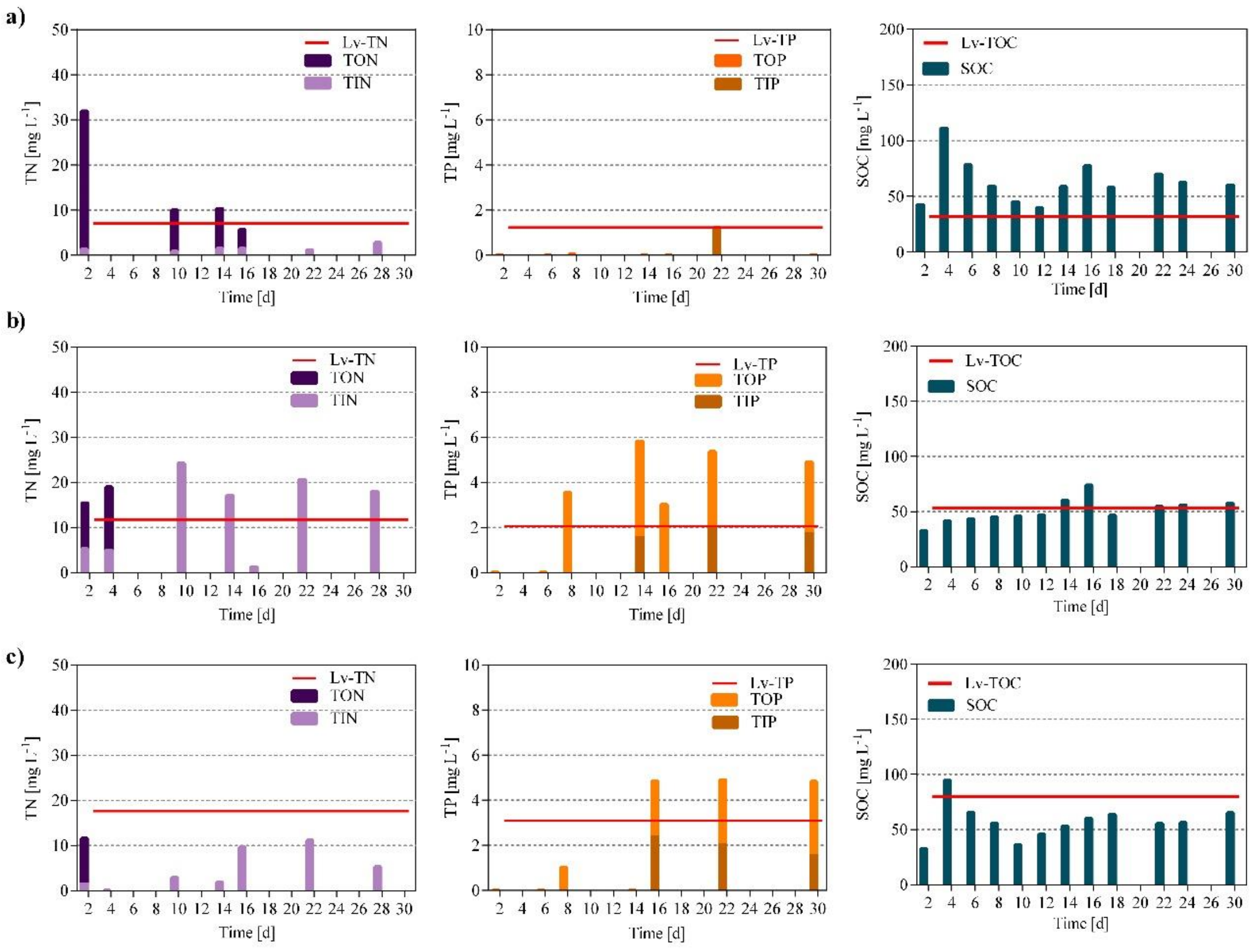

Fig. 7. Time course of effluent concentration for total nitrogen (TIN), total phosphorus (TP) and total organic carbon (TOC) in [mg $L^{-1}$]. a) $SC_{10}$, b) $PSBR_6$ and c) $PSBR_4$. Lines represents influent average volumetric loads in [mg $L^{-1}$ $d^{-1}$].

Table 1. Average (standard deviation) of the main quality parameters of the digestate, secondary effluent and the influent (mixture of digestate and secondary effluent) (*n*=5–10).

| Parameter | Digestate | Secondary effluent | Influent[a] |
|---|---|---|---|
| TSS [g $L^{-1}$] | 28.09 (6.15) | -[b] | 0.56 (0.12) |
| VSS [g $L^{-1}$] | 21.50 (4.45) | -[b] | 0.43 (0.09) |
| TOC [mg $L^{-1}$] | 15999.74 (1337.55) | 18.35 (2.06) | 320.36 (26.79) |
| SOC [mg $L^{-1}$] | 3117.68 (315.51) | 18.37 (2.05) | 62.72 (6.35) |
| TIC [mg $L^{-1}$] | 5538.51 (845.8) | 21.50 (1.78) | 111.19 (16.95) |
| TAN [mg $L^{-1}$] | 1020 (93.95) | -[c] | 20.41 (1.88) |
| $N\text{-}NO_2^-$[mg $L^{-1}$] | <LOD | 1.73 (0.29) | 1.73 (0.29) |
| $N\text{-}NO_3^-$[mg $L^{-1}$] | <LOD | 8.12 (3.07) | 8.12 (3.07) |
| TIN [mg $L^{-1}$] | 1020.02 (257.05) | 9.85 (4.94) | 30.25 (5.24) |
| TN [mg $L^{-1}$] | 3555.50 (1036.51) | 23.50 (1.54) | 71.58 (20.76) |
| TON [mg $L^{-1}$] | 2535.02 (779.45) | 13.65 (3.91) | 40.18 (12.06) |
| TIP [mg $L^{-1}$] | <LOD | 1.97 (3.41) | 1.97 (3.41) |
| TOP [mg $L^{-1}$] | 920 (90) | -[d] | 18.40 (1.80) |
| TP [mg $L^{-1}$] | 1000.8 (93) | 0.69 (0.52) | 20.03 (1.87) |
| TOC:TIC[e] | - | - | 2.88:1 |
| TN:TP[e] | - | - | 7.9:1 |

<LOD Limit of detection.
[a] The influent was prepared as a dilution of digestate within secondary effluent in a 1:50 ratio.
[b] TSS and VSS in the secondary effluent corresponded to values <0.03 g $L^{-1}$.
[c] TAN in the secondary effluent corresponded to values <0.002 mg $L^{-1}$.
[d] TOP in the secondary effluent corresponded to values <0.09 mg $L^{-1}$.
[e] Ratio in molar bases.

Table 2. Average (standard deviation) of the nutrients volumetric loading (Lv) in each photobioreactor.

| Parameter | $SC_{10}$ | $PSBR_6$ | $PSBR_4$ |
|---|---|---|---|
| Lv-TOC [mg $L^{-1}$ $d^{-1}$] | 32.04 (2.68) | 53.31 (4.46) | 80.09 (6.70) |
| Lv-TIC [mg $L^{-1}$ $d^{-1}$] | 11.12 (1.70) | 18.50 (2.82) | 27.80 (4.24) |
| Lv-TAN [mg $L^{-1}$ $d^{-1}$] | 2.04 (0.19) | 3.39 (0.31) | 5.10 (0.47) |
| Lv-N-$NO_3^-$ [mg $L^{-1}$ $d^{-1}$] | 0.81 (0.31) | 1.35 (0.51) | 2.03 (0.77) |
| Lv-N-$NO_2^-$ [mg $L^{-1}$ $d^{-1}$] | 0.17 (0.03) | 0.29 (0.05) | 0.43 (0.07) |
| Lv-TIN [mg $L^{-1}$ $d^{-1}$] | 3.03 (0.52) | 5.03 (0.87) | 7.56 (1.31) |
| Lv-TON [mg $L^{-1}$ $d^{-1}$] | 4.02 (1.21) | 6.69 (2.01) | 10.05 (3.02) |
| Lv-TN [mg $L^{-1}$ $d^{-1}$] | 7.04 (2.08) | 11.72 (3.45) | 17.61 (5.19) |
| Lv-TIP [mg $L^{-1}$ $d^{-1}$] | 0.20 (0.34) | 0.33 (0.57) | 0.49 (0.85) |
| Lv-TOP [mg $L^{-1}$ $d^{-1}$] | 1.04 (0.11) | 1.73 (0.18) | 2.60 (0.26) |
| Lv-TP [mg $L^{-1}$ $d^{-1}$] | 1.24 (0.45) | 2.06 (0.75) | 3.09 (1.12) |